**Title:** Empirical estimates suggest most published medical research is true

**Authors:** Leah R. Jager[1] and Jeffrey T. Leek[2]*

**Affiliations:**
1. Department of Mathematics, United States Naval Academy, Annapolis, MD 21402, USA
2. Department of Biostatistics, Johns Hopkins Bloomberg School of Public Health, Baltimore, MD 21205, USA
* To whom correspondence should be addressed

**Abstract:**

The accuracy of published medical research is critical both for scientists, physicians and patients who rely on these results. But the fundamental belief in the medical literature was called into serious question by a paper suggesting most published medical research is false. Here we adapt estimation methods from the genomics community to the problem of estimating the rate of false positives in the medical literature using reported P-values as the data. We then collect P-values from the abstracts of all 77,430 papers published in *The Lancet, The Journal of the American Medical Association, The New England Journal of Medicine, The British Medical Journal,* and *The American Journal of Epidemiology* between 2000 and 2010. We estimate that the overall rate of false positives among reported results is 14% (s.d. 1%), contrary to previous claims. We also find there is not a significant increase in the estimated rate of reported false positive results over time (0.5% more FP per year, P = 0.18) or with respect to journal submissions (0.1% more FP per 100 submissions, P = 0.48). Statistical analysis must allow for false positives in order to make claims on the basis of noisy data. But our analysis suggests that the medical literature remains a reliable record of scientific progress.

**Introduction:**

Scientific progress depends on the slow, steady accumulation of data and facts about the way the world works. The scientific process is also hierarchical, with each new result predicated on the results that came before.  When developing new experiments and theories, scientists rely on the accuracy of previous discoveries, as laid out in the published literature. The accuracy of published research is even more critical in medicine -- physicians and patients may make treatment decisions on the basis of the latest medical research.

The fundamental belief in the medical literature was called into serious question by a paper suggesting most published medical research is actually false[1]. The claim is based on the assumption that most hypotheses considered by researchers have a low pre-study probability of being successful. The suggested reasons for this low pre-study probability are small sample sizes, bias in hypothesis choice due to financial considerations, or bias due to over testing of hypotheses in "hot" fields. Based on this assumption, many more false hypotheses would be tested than true hypotheses.

The rest of the argument is based on standard properties of hypothesis tests. If all P-values less than some value α are called significant than on average α% of false hypotheses will be reported as significant. The usual choice for α in most medical journals is 0.05, resulting in 5% of false hypotheses being reported as significant. Meanwhile, true hypotheses will be called significant at a much higher rate, β. Many studies are designed so that β is a large value like 0.8; so that 80% of true hypotheses will be called significant.

However, if many more false hypotheses are tested than true hypotheses, the fraction of significant results corresponding to true hypotheses will still be low. The reason is that 5% of a very large number of hypotheses is still larger than 80% of a very small number of hypotheses [2] (**Figure 1**). The argument is plausible, since hypothesis testing is subject to error[3], the pressure to publish positive results is strong [4] [5], and the number of submitted papers is steadily growing (**Figure 2**).

The assumptions that lead to the original result have been called into question, but most counter arguments simply make a different set of assumptions about the rate of false hypotheses being tested [6]. On the other hand, evidence based medicine focuses on determining whether specific medical hypotheses are true or false by aggregating data from multiple studies and performing meta-analyses [7] [8]. But to date, there has been no empirical approach for evaluating the rate of false positives across an entire journal or across multiple journals.

To fill this gap, here we develop a statistical algorithm to directly estimate the proportion of false discoveries in the medical literature based only on reported, significant P-values. Our approach is derived from similar statistical methods for estimating the rate of false positives in genomic[9,10,11,12] or brain-imaging studies[13], where many hypotheses are tested simultaneously. There are serious problems with interpreting individual P-values as evidence for the truth of the null hypothesis[14]. It is also well-established the reporting measures of scientific uncertainty such as confidence intervals are critical for appropriate interpretation of published results[15]. However, there are well-established and statistically sound methods for estimating the rate of false positives among an aggregated set of tested hypotheses using P-values[9-11]. Since P-values are still one of the most commonly reported measures of statistical significance, it is possible to collect these P-values and use them as data to estimate the rate of false positives in the medical literature.

We collect all 5,322 P-values reported in the abstracts of the 77,430 papers published in the *The Lancet, The Journal of the American Medical Association (JAMA), The New England Journal of Medicine (NEJM), The British Medical Journal (BMJ),* and *The American Journal of Epidemiology (AJE)* between the years 2000 and 2010. Based on these data we are able to empirically estimate the rate of false positives in the medical literature and trends in false positive rates over time. We show that despite the theoretical arguments to the contrary, the medical literature remains a reliable record of scientific progress.

**Methods:**

*Estimating the science-wise false discovery rate:*

Our goal is to estimate the rate that published research results are false positives. False positives, for our analysis, are cases where the null hypothesis is true in a hypothesis testing framework. We will call this quantity $\pi_0$. This is the same rate that was considered in the original paper claiming most published results are false [1]. Since we do not know all of the hypotheses tested by all researchers, we cannot estimate the probability that research hypotheses are false. But we can observe the reported P-values in the medical literature.

By definition, the P-values for false positive findings are uniformly distributed between 0 and 1 [16]. The reported P-values in the literature represent a subset of all P-values computed. Specifically, they are most frequently the statistically significant P-values. The usual threshold for significance is P < 0.05, so we focus on developing a method that can be applied to only the reported P-values less than 0.05, since they can be observed. Another statistical property means that if we consider only the P-values that are less than 0.05, the P-values for false positives must be distributed uniformly between 0 and 0.05, denoted *U(0,0.05)*. Each true positive P-value may be drawn from a different distribution, but extensive statistical research has shown that when considered as one group, the distribution of true positive P-values can be accurately modeled as a Beta distribution, *β(a,b)*[17-20]. If we only consider the P-values less than 0.05, the distribution for true positive P-values can be modeled by a Beta distribution truncated at 0.05 *tβ(a,b; 0.05)*. This conditional distribution models the behavior of reported P-values when scientists report all P-values less than 0.05 as significant, including the case where they test many hypotheses and only report the P-values less than 0.05 (see Supplementary Material).

For any specific P-value we do not know whether it represents a true positive or a false positive. So with probability $\pi_0$ it corresponds to the false positive distribution and with probability (1-$\pi_0$) it corresponds to the true positive distribution. So the P-value distribution can be modeled as a mixture of these two distributions [10][11]:

$$f(p|a,b,\pi_0) = \pi_0 U(0,0.05) + (1-\pi_0)t\beta(a,b; 0.05) \quad [1]$$

A similar approach is taken in genomics or brain-imaging studies when many hypotheses are tested simultaneously, although in that case all P-values are observed, not just those less than 0.05. The parameters in equation [1] can be directly estimated using the EM-algorithm [21].

One additional complication to equation [1] is that some P-values are not reported exactly, but are truncated. For example, a P-value of 0.000134343 may be reported as P ≤ 0.01. Since we do not observe the exact value of these P-values, they are censored. This is a similar problem encountered in the analysis of survival data, when patients are lost to follow-up [22]. If we treat the P-values that are reported as truncated (those reported with < or ≤, rather than =, in the abstract) we can apply standard survival analysis models to extend model [1] to handle the censored observations[23]. The key assumption here, as in standard survival analysis, is that censoring is independent of the P-value distributions. This is reasonable because it is unlikely that among all scientists the decision to round is correlated with choice of statistical models across scientists, labs, and journals.

A second complication is that some P-values are rounded, so that a P-value of 0.013 is reported as 0.01. Since P-values are continuous random variables the probability of being exactly a round value like 0.01 is zero. So the observation that many P-values are reported as 0.01, 0.02, 0.03, 0.04, or 0.05 strongly suggests that these P-values are rounded. We model these P-values as multinomial random variables, taking on one of the 5 values 0, 0.01, 0.02, 0.03, 0.04, and 0.05. The probability of each rounded value is equal to the total probability assigned to all the P-values that round to that value. We can calculate these probabilities by integrating the distributions over the intervals rounding to 0, 0.01, 0.02, 0.03, 0.04, and 0.05: [0,0.005), [0.005,0.015), [0.015,0.025), [0.025,0.035), [0.035,0.045), and [0.045,0.05], respectively[24][25]. Again, the assumption is that rounding is independent of P-value, which is again likely because it would only happen if there is a correlation between the choice to round and the choice of statistical methods and analyses across scientists, labs, and journals.

With these modifications in place, we can now use the EM-algorithm to estimate the parameters in the model. Specifically, we can estimate $\pi_0$, the rate of false positives. We can apply our algorithm to the P-values from all journals and all years to estimate an overall rate of false positives, or we can apply our algorithm individually to specific journals or years to estimate journal and year specific false positive rates. Full mathematical details of our approach and R code for implementing our models are available in the Supplemental Material.

*Collecting P-values from medical publications*

We wrote a computer program in the R statistical programming language (http://www.R-project.org/) to collect the abstracts of all papers published in *The Lancet, The Journal of the American Medical Association (JAMA), The New England Journal of Medicine (NEJM), The British Medical Journal (BMJ),* and *The American Journal of Epidemiology (AJE)* between 2000 and 2010. Our program then parsed the text of these abstracts to identify all instances of the phrases "P =", "P <", "P ≤", allowing for a space or no space between "P" and the comparison symbols. Our program then extracted both the comparison symbol and the numeric symbol following the comparison symbol. We scraped all reported P-values in abstracts, independent of study type. The P-values were scraped from http://www.ncbi.nlm.nih.gov/pubmed/ on January 24, 2012. A few manual changes were performed to correct errors in the reported P-values due to variations in the reporting of scientific notation as detailed in the R code. To validate our procedure, we selected a random sample (using the random number generator in R) of abstracts and compared our collected P-values to the observed P-values manually. The exact R code used for scraping and sampling and the validated abstracts are available in the Supplemental Material.

*Obtaining Journal Submission Data*

We also directly contacted the editors of these journals, and they supplied data on the number of submitted manuscripts to their respective journals. We were only able to obtain publication data for the years 2006-2010 for the *Lancet* and for the years 2003-2010 for *BMJ*.

*False positive rates over time*

We used a linear mixed model to estimate the rate of increase or decrease in false positive rates over time [26]. The dependent variable was the estimated false positive rate for a journal in each year. The independent variable was the year. We also fit a linear mixed model relating the false positive rate to the number of submissions for each journal. Each model included a journal-specific random effect.

*Reproducible Research*

A key component of computational research is that the results be reproducible. We have adhered to the standards of reproducible research by both outlining all of our statistical methods in the supplemental material and making the exact R code used to produce our results available at https://github.com/jtleek/swfdr. By distributing our methods, code, and software our approach adheres to the highest standards of reproducible computational research [27].

**Results:**

Our computer program collected the abstracts from all 77,430 papers published in *The Lancet, The Journal of the American Medical Association (JAMA), The New England Journal of Medicine (NEJM), The British Medical Journal (BMJ),* and *The American Journal of Epidemiology (AJE)* between the years 2000 and 2010. The abstracts were mined for P-values, as described in the Methods section. Of the mined abstracts, 5,322 reported at least one P-value according to our definition. The relatively low rate of P-values in abstracts is due to (1) many articles not reporting P-values in the abstracts since they are essays, review articles or letters and (2) a trend toward decreased reporting of P-values and increased reporting of other statistical measures such as estimates and confidence intervals. Most reported P-values were less than the usual significance threshold of 0.05: 80% for *The Lancet,* 76% for *JAMA*, 79% for *NEJM*, 76% for *BMJ*, and 75% for *AJE*. Among the papers that reported P-values, a median of 2 P-values were reported (m.a.d. 1.5). Our validation analysis of 10 randomly selected abstracts showed that we correctly collected 20/21 P-values among these abstracts, with no false positive P-values (Supplemental Material).

The distributions of collected P-values showed similar behavior across journals and years (**Figure 3**). Most P-values were substantially less than 0.05, with spikes at the round values 0.01, 0.02, 0.03, 0.04, and 0.05. There was some variation across journals, likely due to variation in editorial policies and types of manuscripts across the journals. We applied our algorithm to all the P-values from all journals and all years to estimate the overall rate of false positives in the medical literature. The estimated rate of false positives among published results was 14% (s.d. 1%). We did find variation among journals with false positive rates of 19% (s.d. 3%) for *Lancet*, 17% (s.d. 2%) for *JAMA*, 11% (s.d. 2%) for *NEJM*, and 17% (4%) for *BMJ.* To compare with the four major medical journals, we calculated an estimate of the rate of false positives for the *American Journal of Epidemiology (AJE)*. The estimated rate was 11% (s.d. 4%), similar to the *NEJM*. The *AJE* is an epidemiological journal and publishes substantially different types of studies than the medical journals. Specifically, there is a substantially greater focus on observational studies. This suggests that the false positive rate is somewhat consistent across different journal and study types.

Next we considered the rate of false positives over time. We applied our algorithm to estimate the rate of false positives separately for each journal and each year (**Figure 4**). We fit a linear mixed model including a journal specific random effect as detailed in the methods and there was no significant trend in false positive rates for any journal over time (0.5% more FP per year, P = 0.18). Similarly, the positive trend in submission rates for these journals (**Figure 2**) was not associated with an increase in false positive rates over the period 2000-2010. (0.1% more FP per 100 submissions, P = 0.48).

**Discussion:**

Here we proposed a new statistical method for estimating the rate of false positives in the medical literature directly based on reported P-values. We then collected data from the major medical journals and showed that most published medical research results in these journals are not false. A similar result held for the epidemiological journal *American Journal of Epidemiology*, even though this journal publishes a substantially different type of result in general. Our results suggest that while there is an inflation of false positive results above the nominal 5% level [3], but the relatively minor inflation in error rates does not merit the claim that most published research is false. Although our results disprove the original claim that most published research is false, they do not invalidate the criticisms of standalone hypothesis tests for statistical significance that were raised. Specifically, it is still important to report estimates and confidence intervals in addition to or instead of P-values when possible so that both statistical and scientific significance can be judged by readers [15,28]. In some cases, a true positive hypothesis may still have a very small effect size and still lead to a small P-value if the sample size is large enough, although these cases may not be scientifically interesting.

An important consideration is that we have focused our analysis on the major medical journals. An interesting avenue for future research would be to consider more specialized medical journals to determine journal characteristics that associate with variation in the false positive rate. Another limitation of our study is that we consider only P-values in abstracts of the papers, which generally correspond to primary results. Another potentially informative analysis could focus on breaking down the rate of false positives by result type. Despite these limitations we have performed the first global empirical analysis of the rate of false positives in the major medical journals and we have shown that theoretical claims to the contrary, the data show that the medical and scientific literature remain a reliable record of scientific progress.

**Acknowledgements:**

JTL was partially funded by a JHSPH Faculty Innovation Fund Award.

**Supplementary Material:**

Supplementary material is available from: http://biostat.jhsph.edu/~jleek/swfdr_supp.pdf. Code to reproduce our analyses is available from: https://github.com/jtleek/swfdr.

**Figures:**

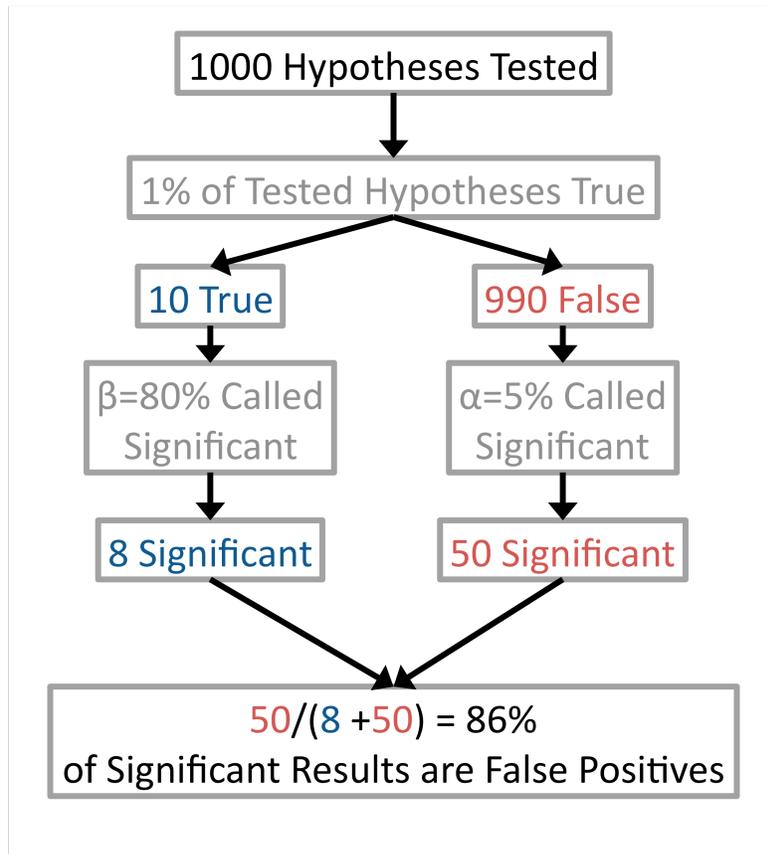

**Figure 1. The theoretical argument suggesting most published research is false.** If the probability a research hypothesis is true is low then most tested hypotheses will be false. The definition of P-values and customary significance cutoffs mean that α% of false positive hypotheses and β% of true positive hypotheses will be called significant. If only 1% of tested hypotheses are true and the customary values of α=5% and β=80% are used, then 86% of reported significant results will be false positives. This final percent of published results corresponding to false positives is the quantity we estimate. A version of this figure appeared on the blog Marginal Revolution and is reproduced with permission[2].

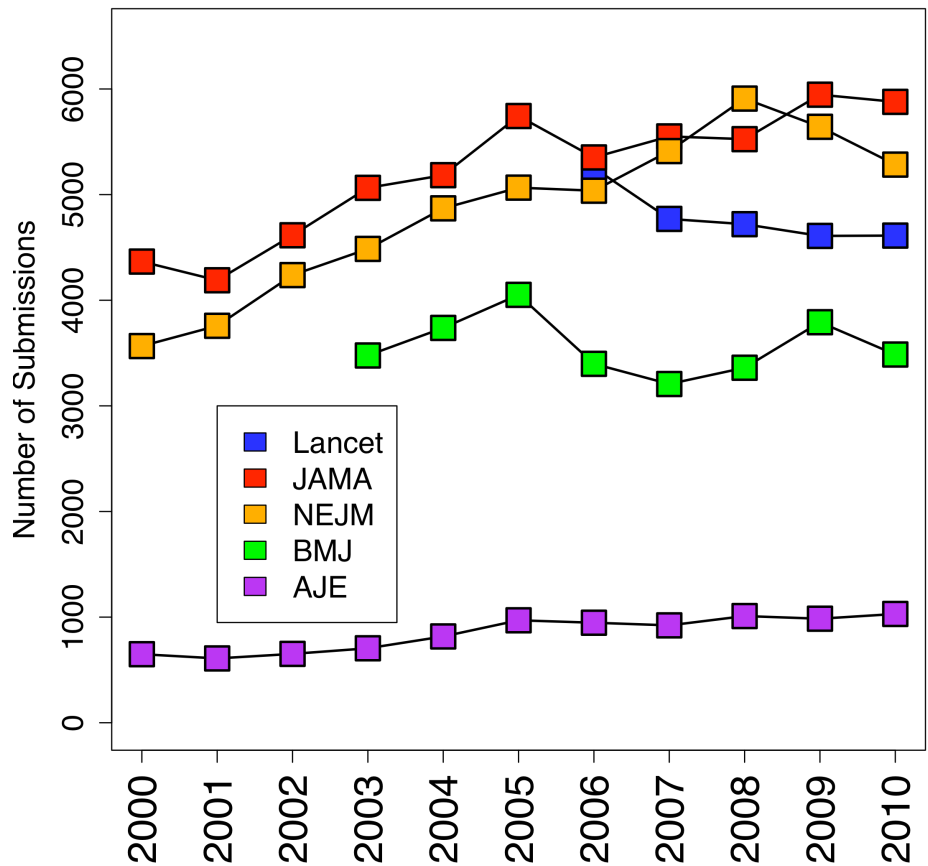

**Figure 2. Major medical journal submissions are increasing over time.** A plot of the number of submissions to the major medical journals *The Lancet, The Journal of the American Medical Association (JAMA), The New England Journal of Medicine (NEJM), The British Medical Journal (BMJ)* and the flagship epidemiological journal *The American Journal of Epidemiology (AJE)* for the the years 2000-2010. Submission data is only available for the years 2006-2010 for *The Lancet* and the years 2003-2010 for *The BMJ*.

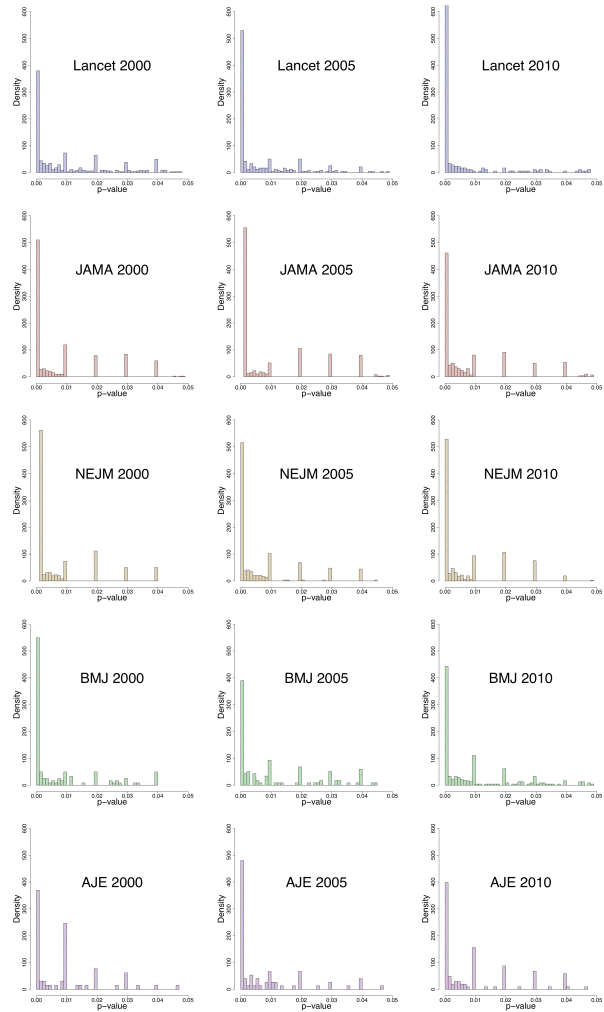

**Figure 3. Histogram of P-values < 0.05**. The observed P-value distributions for all P < 0.05 scraped from Pubmed for the journals *The American Journal of Epidemiology*, *The Journal of the American Medical Association*, *The New England Journal of Medicine*, *The British Medical Journal*, and *The Lancet* in the years 2000, 2005, and 2010

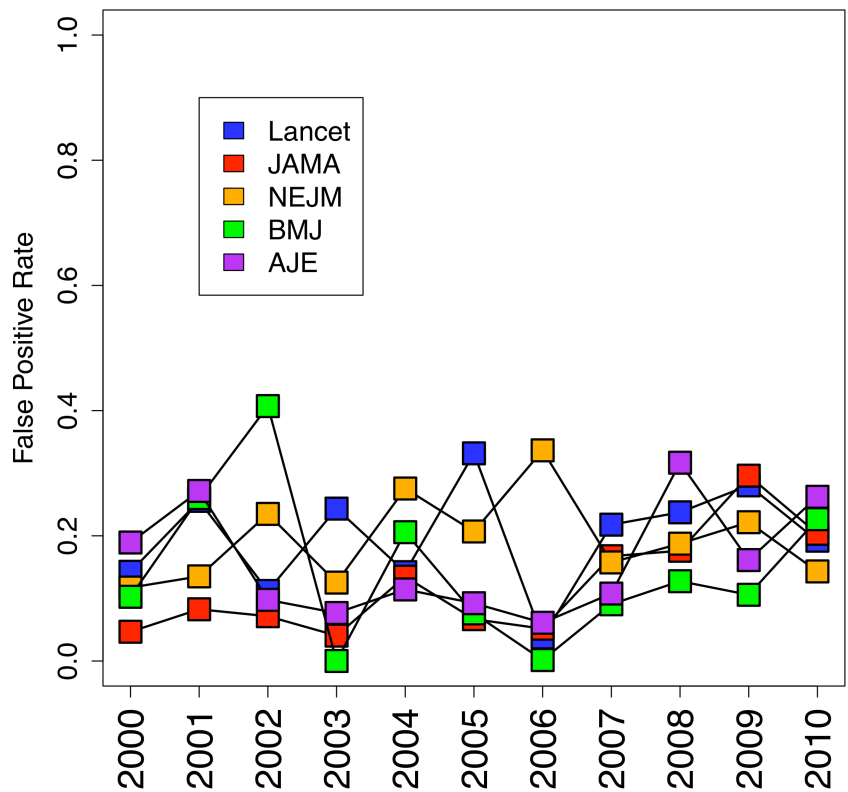

**Figure 4. Estimated false positive rates for the years 2000-2010 by journal.** The estimated false positive rates for the the journals *The American Journal of Epidemiology*, *The Journal of the American Medical Association*, *The New England Journal of Medicine*, *The British Medical Journal*, and *The Lancet* in the years 2000, 2005, and 2010. There is no significant trend in false positive rates over time or with increasing numbers of submissions.